\documentstyle[12pt,epsfig]{article}

\newcommand{\la}[1]{\label{#1}}

\newlength{\numlen}

\newcommand{\cen}[1]{\multicolumn{1}{c}{#1}}

\newlength{\indexlength}

\newcommand{\be}{\begin{equation}}
\newcommand{\ee}{\end{equation}}
\newcommand{\ba}{\begin{eqnarray}}
\newcommand{\ea}{\end{eqnarray}}

\newcommand{\rmi}[1]{{\mbox{\scriptsize #1}}}

\newcommand{\eq}{eq.~}

\newcommand{\nr}[1]{(\ref{#1})}

\newcommand{\tr}{\mbox{Tr\,}}

\newcommand{\bfk}{\mbox{\bf k}}
\newcommand{\bfp}{\mbox{\bf p}}
\newcommand{\bfx}{\mbox{\bf x}}

\newcommand{\half}{{\scriptstyle{1\over2}}}

\newcommand{\fr}[2]{{\frac{#1}{#2}}}

\begin{document}

\begin{titlepage}

\hfill CERN-TH.6901/93\\

\begin{centering}
\vfill

{\bf A LATTICE MONTE CARLO STUDY OF THE HOT ELECTROWEAK PHASE
TRANSITION}
\vspace{1cm}

K. Kajantie$^{a,b}$, K. Rummukainen$^a$ and M.
Shaposhnikov$^a$\footnote{On leave of absence from Institute for
Nuclear Research of Russian Academy of Sciences, Moscow 117312,
Russia} \\

\vspace{1cm}
{\em $^a$Theory Division, CERN,\\ CH-1211 Geneva 23, Switzerland}
\vspace{0.3cm}

{\em $^b$Department of Theoretical Physics,
P.O.Box 9, 00014 University of Helsinki, Finland\\}
\vspace{2cm}
{\bf Abstract}

\end{centering}

\vspace{0.3cm}\noindent

We study the finite temperature electroweak phase transition with
lattice perturbation theory and Monte Carlo techniques. Dimensional
reduction is used to approximate the full four-dimensional SU(2) + a
fundamental doublet Higgs theory by an effective three-dimensional
SU(2) + adjoint Higgs + fundamental Higgs theory with coefficients
depending on temperature via screening masses and mass counterterms.
Fermions contribute to the effective theory only via the $N_F$ and
$m_{\rm top}$ dependence of the coefficients. For sufficiently small
lattices ($N^3 < 30^3$ for $m_H$ = 35 GeV) the study of the one-loop
lattice effective potential shows the existence of the {\em second}
order phase transition even for the small Higgs masses. At the same
time, a clear signal of a {\em first order} phase transition is seen
on the lattice simulations with a transition temperature close to but
less than the value determined from the perturbative calculations.
This indicates that the dynamics of the first order electroweak phase
transition depends strongly on non-perturbative effects and is not
exclusively related to the so-called $\phi^3$ term in the effective
potential.

\vfill \vfill
\noindent

CERN-TH.6901/93\\
May 1993
\end{titlepage}

\section{Introduction}
The interest in the study of the high temperature phase transitions in
gauge theories, initiated a long time ago in \cite{kir}, has been
revived now in connection with possible generation of the baryonic
asymmetry of the universe at the electroweak scale (see, e.g., reviews
\cite{s:rev,far} and references therein). The common belief, based on
the perturbative calculations of the effective potential for the
scalar field, is that the electroweak phase transition for moderate
Higgs boson masses is weakly first order (i.e. temperature
metastability range is small compared with the critical temperature).
An incomplete list of references containing perturbative analysis of
the phase transitions is contained in \cite{linde}-\cite{buch}. All
perturbative calculations, however, work only for sufficiently large
values of the scalar field, namely $\phi \gg g T$. This estimate
arises as follows. Roughly speaking, there are at least three relevant
mass scales at high temperatures. The first one is related to the
Debye screening of the gauge charge and is of the order of $m_D =C_D g
T$ (for the electroweak theory with $N_F=6$ $C_D = \frac{11}{6}$). The
second one is associated with the non-abelian magnetic sector of gauge
theories, $m_M = C_M g^2 T$ ($C_M$ is some number yet to be
determined) and the third one with the gauge boson mass induced by the
Higgs mechanism, $m_W = \frac{1}{2} g \phi$ with $\phi$ being the
condensate of the scalar field. The naive loop expansion for the
effective potential works well provided $m_W \gg m_D$, i.e. for $\phi
\gg T$. This condition is usually not satisfied for the interesting
range of fields and temperatures. The improved loop expansion which
takes into account Debye mass effects in all orders of perturbation
theory works for a larger interval of the scalar condensate. However,
due to the infrared problem in the high temperature dynamics of gauge
theories \cite{linde:pl80} any of the loop expansions break down for
$m_W < m_M$ and $\phi < 2 C_M g T$. Due to the fact that the coupling
constant in electroweak theory is not so small ($g \approx 2/3$) it is
important to know the numerical value of the coefficient $C_M$. If it
is small, then perturbative calculations of the effective potential
are valid in an interesting range of parameters, while a large value
of $C_M$ would imply the lack of any knowledge of the dynamics of the
phase transition.

The appearance of the magnetic mass is a purely non-perturbative
phenomenon and not a lot is known about its value. It is associated
with the confinement scale of 3-dimensional gauge theory derived by
dimensional reduction from 4-dimensional theory at high temperatures.
Different correlation lengths in 3-dimensional SU(2) gauge theory were
studied by lattice Monte-Carlo methods in \cite{kks:magn}. It was
found there that the 3-dimensional $0^{++}$ glueball mass (inverse
screening length in this channel) is about $\sim 2 g^2 T$. This
indicates that non-perturbative effects are rather large and that
presumably $C_M \sim 2$. If true, a perturbative analysis of the
effective potential in a parameter range relevant for cosmology has
low chances to remain if force after non-perturbative effects are
taken into account.

In order to clarify this question, one has to use non-perturbative
methods for the study of the effective potential. They are provided by
lattice Monte-Carlo simulations. Some preliminary data derived with
the use of 4-dimensional lattices is already available
\cite{heller}-\cite{bunk}.  There is an indication \cite{bunk} that
the strength of the first order phase transition is enhanced in
comparison with perturbative calculations.

The purpose of this paper is to combine perturbative and
non-perturbative methods for the study of the finite $T$ electroweak
transition. Namely, with the use of dimensional
reduction~\cite{nadkarni}-\cite{lacock} one can first derive a
3-dimensional effective action with perturbatively calculable
coefficients, summing up thus the effects of the Debye screening which
are well understood. Then, one can simulate the phase transition with
lattice Monte-Carlo methods.

There are a number of advantages of this method in comparison with
4-dimensional lattice simulations. First, it separates the physics in
which we are reasonably confident (Debye screening) from the unknown
non-perturbative 3-dimensional physics. In other words, any
4-dimensional simulations contain perturbative noise which has nothing
to do with non-perturbative effects determining the order of the phase
transition. Moreover, perturbation theory signals that the first order
nature of the electroweak phase transition is an exclusively
3-dimensional phenomenon, since the so-called $\phi^3$ term in the
effective potential (this term induces a jump of the order parameter)
arises due to infrared singularities in loop integrations at high
temperatures. The last, but not least advantage is that a lattice
simulation of 3-dimensional theories is less time consuming and more
transparent from the point of view of scaling behaviour.

The paper is organized as follows. In Section 2 we derive the
3-dimensional effective action for our system. In Section 3 we compute
the effective potential in the continuum 3-dimensional theory and show
how the first order nature of the phase transition arises in the loop
expansion of the effective potential. In Section 4 we study the
effective potential for the lattice version of the theory which is
then compared in Section 5 with Monte Carlo simulations. Section 6
contains our conclusions.

\section{The effective action}

Finite $T$ field theoretic systems are characterised by fields defined
over the interval $0<\tau<\beta\equiv 1/T$ in imaginary time and
extended beyond this region by the condition of periodicity
(antiperiodicity) with the period $\beta$ for bosonic (fermionic)
fields. If the action is expressed in terms of the Fourier components,
the quadratic terms are of the type $[(2\pi nT)^2 +
\bfk^2]|A(n,\bfk)|^2$, where $A(n,\bfk)$ is a generic bosonic field
and $n=-\infty,\dots,+\infty$. At high $T$ and $k<2\pi T$ the
nonstatic modes $A(n\not=0,\bfk)$ are thus suppressed by the factor
$(2\pi nT)^2$ relative to the static $A(0,\bfk)$ modes. The idea of
dimensional reduction is to exploit this suppression by integrating
over the nonstatic modes in $S[A(n=0,\bfk),A(n\not=0,\bfk)]$ and
deriving in this way an effective action $S_\rmi{eff}[A(0,\bfk)]$ for
the dominant static modes. We shall below carry this out for the
four-dimensional SU(2) + a fundamental Higgs theory, taking
$g^\prime=0$ in the electroweak sector of the standard model.  The
effective theory then is a three-dimensional SU(2) + a fundamental
Higgs + an adjoint Higgs model with well defined $T$-dependent
coefficients.

For fermionic fields the square of the inverse propagator is
$[(2n+1)\pi T]^2 +\bfk^2$ and all modes are suppressed at large $T$.
However, the fermionic fields will enter by changing the coefficients
of the effective action of the bosonic static modes. Their effect can
thus also be studied in this framework. However, as our calculations
so far are only on the one loop level, we can only include a top quark
of mass less than 79 GeV, given by one loop
stability~\cite{onelooptop}.

The starting point is the action of the 4d SU(2) + fundamental Higgs
model
\be
  S[A_\mu^a(\tau,\bfx),\phi_i(\tau,\bfx)]=\int_0^\beta d\tau\int
  d^3x\{ {\scriptstyle{1\over4}} F_{\mu\nu}^aF_{\mu\nu}^a +
  (D_\mu\phi)^\dagger (D_\mu\phi) +\mu^2 \phi^\dagger\phi + \lambda
  (\phi^\dagger\phi)^2\}, \la{4daction}
\ee
in standard notation.  Possible fermionic terms are not shown
explicitly.

Including 1-loop corrections and terms not damped by powers of $1/T$
the dimensionally reduced effective action becomes
\begin{eqnarray}
&&S_{\rmi{eff}}[A_i^a(\bfx),A_0^a(\bfx),\phi_i(\bfx)]
= {1\over T} \int d^3x \biggl\{{1\over4} F_{ij}^aF_{ij}^a +
\fr12 (D_iA_0)^a(D_iA_0)^a + (D_i\phi)^\dagger(D_i\phi)+ \nonumber
\\&&
+\fr12
\biggl[\fr23(1+\fr14+\fr{N_F}4)g^2T^2 -(4+1)g^2T\Sigma_c\biggr]
A_0^aA_0^a +{g^4\over12\pi^2}(1+{1\over16}-{N_F\over8})(A_0^aA_0^a)^2
+
\nonumber \\&&
+\biggl[-\fr12 m_H^2 +({1\over8}g^2+{1\over16}g^2 +\fr12\lambda
+{g^2m_{\rmi{top}}^2\over 8m_W^2})T^2
-(\fr32 g^2 + {3\over4} g^2 +
6\lambda)T\Sigma_c\biggr]\phi^\dagger\phi
+\lambda(\phi^\dagger\phi)^2 + \nonumber \\&&
+\fr14 g^2 A_0^aA_0^a \phi^\dagger\phi \biggr\},\la{3daction}
\end{eqnarray}
where $\Sigma_c$ is the integral
\be
\Sigma_c= \int{d^3p\over (2\pi)^3\bfp^2} \la{Sigmacont}
\ee
depending linearly on the cutoff. In perturbation theory it cancels
against 1-loop divergences, as shown explicitly below. If one wants to
study the system nonperturbatively with a finite cutoff it must be
included for correct continuum limit \cite{parisi}.  We shall fix the
parameters of the effective action by taking $g=2m_W(\sqrt2
G_F)^{-1/2}=2/3$, $m_W=$ 80.6 GeV and giving the value of $m_H$. Then
$\lambda/g^2=m_H^2/(8m_W^2)$.

The effective theory thus is a 3d SU(2) + fundamental Higgs + adjoint
Higgs theory with coefficients depending on $T$, $N_F$, the coupling
constants and the cutoff. All the three kinetic terms, the two $T=0$
potential terms for $\phi$ and the last $A_0-\phi$ coupling term arise
from \eq\nr{4daction} by naive dimensional reduction (taking fields
constant in $\tau$).  The general structure of the 1-loop quadratic
potential terms for $A_0$ and $\phi$ is $c_1T^2-c_2T\times$ cutoff,
where the first term is the usual 4d 1-loop screening mass and the
second term arises from the exclusion of the $n=0$ term in the 1-loop
integral (since one must only integrate over the nonstatic modes).
Equivalently, this term is the mass counterterm of the
superrenormalisable 3d theory. The contributions of the various fields
to these terms in \eq\nr{3daction} are ordered so that in the
coefficient of $\phi^\dagger\phi$ first come the terms with $A_i^a$ in
the loop, then $A_0$, then $\phi$ and last the fermion loop (only in
the $T^2$ terms). In the quadratic and quartic $A_0$ terms the
ordering is $A_i^a,\phi$, fermions in the loop. The $A_0$ loop gives a
contribution of order $g^4$ and is neglected here.  Also neglected are
small 1-loop corrections to the quartic $\phi$ term.

In addition to the mass counterterm \eq\nr{Sigmacont} there in 3d
actually also is \cite{parisi} a logarithmic 2-loop counterterm
$\sim\lambda^2T^2/(16\pi^2)\int dp/|\bfp|$. This is numerically
negligible in our case.

More generally, \eq\nr{3daction} contains terms with higher powers of
fields and their derivatives, allowed by the gauge (BRS) invariance of
the 3d theory. For example, the term $g^6(A_0^aA_0^a)^3/T^2$ would
appear. If we take the correlation length of the $A_0$ field to be
$1/(gT)$ we can estimate from the quadratic term that
$<A_o^aA_0^a>\sim gT^2$.  Thus the ratio of terms with consecutive
powers of $A_0^aA_0^a$ is ${\cal O}(g^3) \ll1$.

\section{The one-loop effective potential in continuum.}

To study the theory defined by the action \eq\nr{3daction} in
perturbation theory, we compute the 1-loop correction
$V_{\rmi{1loop}}(\phi,A_0)$ to the tree potential defined by
\eq\nr{3daction}. One finds before regularisation that
\begin{eqnarray}
&& V_{\rmi{1loop}}(\phi,A_0) =T\int{d^3p\over(2\pi)^3} \biggl\{
2\log(\bfp^2+g^2A_0^2+\fr14 g^2\phi^2)+\log(\bfp^2+\fr14 g^2\phi^2) +
\nonumber \\
&&+\fr32\log(\bfp^2+\mu^2+\lambda\phi^2+\fr14 g^2A_0^2) +\log(\bfp^2
+m_D^2+\fr14g^2\phi^2+\lambda_A A_0^2) +\\
&&+\fr12 \log\{\bfp^4+\bfp^2[\mu^2+m_A^2+(3\lambda+\fr14g^2)\phi^2+
(\fr14g^2+3\lambda_A)A_0^2]+\nonumber \\
&&\qquad+(\mu^2+3\lambda\phi^2+\fr14g^2A_0^2)(m_D^2+\fr14g^2\phi
^2+3\lambda_A A_0^2)
-\fr14g^4A_0^2\phi^2\}\biggr\}, \nonumber
\la{1looppot}
\end{eqnarray}
where $\phi^2=2\phi^\dagger\phi$ and $A_0^2=A_0^aA_0^a$. The first two
terms come from the $A_i$ loop and the remaining ones from the coupled
$A_0,\phi$ loops. Using
\be
  \int{d^3p\over (2\pi)^3} \log[\bfp^2+\mu^2+m^2(\phi)]
  =m^2(\phi)\Sigma_c- {1\over6\pi}[\mu^2+m^2(\phi)]^{3/2} \la{regular}
\ee
one finds that the $\Sigma_c$ terms in \eq\nr{3daction} and
\eq\nr{1looppot} cancel (taking into account the $A_0$ loop term
$-\fr12 5\lambda_AT\Sigma_cA_0^2$ neglected in \eq\nr{3daction} for
smallness).  The finite terms give the quantum correction to the tree
potential.  The resulting 1-loop improved potential is then
\begin{eqnarray}
V(\phi,A_0) &&= \fr12 \mu^2\phi^2 +\fr12 m_D^2 A_0^2
+\fr14\lambda(\phi^2)^2+\fr14\lambda_A(A_0^2)^2+
\fr18 g^2 A_0^2\phi^2 -\nonumber \\
&& -{T\over6\pi} \biggl\{2g^3(\fr14\phi^2+A_0^2)^{3/2}
+\fr18 g^3\phi^3 +\\
&&+(m_D^2+\fr14g^2\phi^2+\lambda_AA_0^2)^{3/2}+
\fr12(m_D^2+\fr14g^2\phi^2+3\lambda_AA_0^2)^{3/2}+ \nonumber \\
&&+\fr32(\mu^2+\lambda\phi^2+\fr14g^2A_0^2)^{3/2}+
\fr12(\mu^2+3\lambda\phi^2+\fr14g^2A_0^2)^{3/2}\biggr\}, \nonumber
\la{pottot}
\end{eqnarray}
where
\be
\mu^2=\gamma(T^2-T_0^2),\quad \gamma=\fr3{16}g^2+\fr12\lambda+
{g^2m_{\rmi{top}}^2\over 8m_W^2}. \la{defofmu}
\ee
We have, for completeness included the small $\lambda_A$ term,
which will again be neglected from now on.

Studying the minima of $V(\phi,A_0)$ one sees that they can be driven
to nonzero values of $A_0$ only if the negative term is large.  This
demands that at least $g>\pi\sqrt{8/3}$ which is beyond the domain of
validity of this calculation. We conclude that the minimum is always
at $A_0=0$; no condensate is formed. Basically this is due to the
large value of $m_D$ and the small value of the correction, $\sim
-T/6\pi$.

For $A_0 =0$ and $\lambda \ll g^2$ the corrections to the effective
potential have the form
\be
\delta V = -\frac{g^3}{16\pi}T \phi^3 - \frac{T}{4\pi}(m_A^2 +
\frac{1}{4}g^2 \phi^2)^{\frac{3}{2}}.
\ee
It is the first term which gives rise to the first order phase
transition in perturbation theory. Note that the effective
3-dimensional theory correctly takes into account Debye screening,
which decreases the magnitude of the cubic term relative to the naive
loop expansion by the factor $2/3$ \cite{lindeandson}. The second term
corrects the $\gamma$ in \eq\nr{defofmu} by the term
$(-3/16\pi)\sqrt{2/3}g^3$. With this input the equation
\be
{m_H^2\over4T_0^2}={g^2\over2}\biggl(
\fr3{16}+ \fr12 {m_H^2\over8m_W^2} + {m_\rmi{top}^2\over8m_W^2} -
\frac{\sqrt3}{8 \pi\sqrt2}g\biggr)
={m_H^2\over4T_c^2}+{g^4\over32\pi^2}{m_W^2\over m_H^2}
\la{Tpert}
\ee
gives the perturbative result for the transition temperature $T_c$ and
the lower end $T_0<T_c$ of the metastability range (the high $T$ phase
does not exist for $T\le T_0$).  Numerical values for these as well as
for some other relevant quantities as calculated from 1-loop
perturbation theory \cite{anderson}, \cite{enqvist} are given in Table
1.  For $g=2/3$ the relevant formulas are $\xi(T_c)T_c= 9\pi m_H/m_W$,
$T_c/m_W(T_c)=9\pi m_H^2/(2m_W^2)$,
$\sigma/T_c^3=2m_W^5/(243\pi^3m_H^5)$, $L/T_c^4 =
2m_W^4[1/6+m_H^2/(18m_W^2)-m_W^2/(18\pi^2m_H^2)]/(9\pi^2m_H^4)$.
\begin{table}[h]
\center
\begin{tabular}{ccrllrll}\hline
  \cen{$m_H$} & \cen{$T_0$} & \cen{$T_c$} & \cen{$T_+$} &
\cen{$\xi(T_c)$}& \cen{$m_W^{-1}(T_c)$} &
  \cen{$\sigma$} & \cen{$L$} \\ \hline
35 & 90.89 &   95.24 & 95.83 & 12.28/$T_c$ &2.67/$T_c$ & 0.017$T_c^3$ &
0.085$T_c^4$ \\
80 & 182.34&   183.55 & 183.71 & 28.1/$T_c$ &13.9/$T_c$ & 0.00028$T_c^3$ &
0.0044$T_c^4$ \\
\hline
\end{tabular}
\caption[1]{Values of $T_c$, the lower ($T_0$) and upper ($T_+$) ends
  of the metastability range, the correlation length $\xi$ for the
  Higgs field at $T=T_c$, the gauge field correlation length
  $1/m_W(T_c)$, the interface tension $\sigma$ and the latent heat $L$
  as calculated from 1-loop perturbation theory for $m_H=$ 35 and 80
  GeV, for $g=2/3$ and for no fermions included.  \la{table1}}
\end{table}

\section{The lattice action and effective potential.}
\subsection{Lattice action.}

To latticize \eq\nr{3daction} we go over to the matrix representation
$A_0=A_0^aT^a$, $T^a=\half\sigma_a$, $\phi\to\Phi=
(\phi_0+i\sigma_i\phi_i)/\sqrt2$ and rescale the fields by
\be
igaA_0\to A_0,\qquad\qquad \Phi\to \sqrt{{T\over
a}{\beta_H\over2}}\Phi,
\la{rescaling}
\ee
where $a$ is the lattice spacing.
The lattice action on an $N^3$ lattice then becomes
\begin{eqnarray}
&&S= \beta_G \sum_x \sum_{i<j}(1-\fr12 \tr P_{ij}) +\nonumber \\
&&+ \fr12\beta_G  \sum_x \sum_i[\tr
A_0(\bfx)U_i^{-1}(\bfx)A_0(\bfx+i)
U_i(\bfx) - \tr A_0^2(\bfx)]+
\nonumber \\
&& + \sum_x \biggl\{10\Sigma(N^3) - {5+N_F\over3}
{4\over g^2\beta_G} \biggr\} \fr12 \tr A_0^2(\bfx) + \nonumber \\
&& + \sum_x {g^2\beta_G\over 3\pi^2} (1+\fr1{16} -{N_F\over8})
( \fr12 \tr A_0^2(\bfx) )^2 + \la{latticeaction}\\
&& + \beta_H \sum_x \sum_i\bigl[ \fr12\tr\Phi^\dagger(\bfx)\Phi(\bfx)
- \fr12\tr\Phi^\dagger(\bfx)U_i(\bfx)\Phi(\bfx+i) \bigr]+\nonumber \\
&&+ \sum_x \bigl[ (1-2\beta_R - 3\beta_H)
\fr12\tr\Phi^\dagger(\bfx)\Phi(\bfx) + \beta_R
 \bigl( \fr12\tr\Phi^\dagger(\bfx)\Phi(\bfx) \bigr)^2 \bigr]
+\nonumber \\
&& - \fr12\beta_H \sum_x
\bigl[\fr12\tr A_0^2(\bfx) \fr12\tr\Phi^\dagger(\bfx)\Phi(\bfx)
\bigr],\nonumber
\end{eqnarray}
where
\be
\Sigma(N^3)={1\over 4N^3}\sum_{n_i=0}^{L-1} {1\over
\sin^2(\pi n_1/N)+\sin^2(\pi n_2/N)+\sin^2(\pi n_3/N)}
\la{Sigmalatt}
\ee
(the term with all $n_i=0$ is omitted;
numerically $\Sigma(N^3)$ =
0.224605, 0.233942, 0.238633, 0.243326, 0.252731 for $N$ =
8,12,16,24,$\infty$, respectively) and
\begin{eqnarray}
&&\beta_G = {4\over g^2}{1\over Ta} \nonumber\\
&&\beta_R={1\over4}\lambda Ta \beta_H^2 =
{m_H^2\over 8m_W^2}{\beta_H^2\over\beta_G} \la{betas}\\
&&\mu^2(T)={2(1-2\beta_R-3\beta_H) \over \beta_H a^2}. \nonumber
\end{eqnarray}
In \eq\nr{betas} $\mu^2(T)$ is the coefficient of the $\phi^\dagger\phi$
term in \eq\nr{3daction} with the continuum $\Sigma_c$ replaced by
$\Sigma(N^3)/a$.
Introducing this relates $\beta_H$ and $T$ for given
values of $g,m_W,m_H$ (and possibly of $m_\rmi{top}$):
\be
{m_H^2 \over4T^2}=\biggl({g^2\beta_G\over4}\biggr)^2
\biggl[ 3-{1\over\beta_H} + 2{m_H^2\over8m_W^2} {\beta_H\over\beta_G}
-{9\over2\beta_G}(1+{m_H^2\over 3m_W^2})\Sigma(N^3) \biggr]
+{g^2\over2}\biggl(
\fr3{16}+ \fr12 {m_H^2\over8m_W^2} +
{m_\rmi{top}^2\over8m_W^2}\biggr), \la{betahT}
\ee

The curve $T=T(\beta_H)$ is plotted in Fig.1 for a few
parameter values. The curves shift to the right (left) with
increasing
$m_H$ and $\Sigma(N^3)$ ($\beta_G$). For example, the values of
$\beta_H$
corresponding to $T=T_0$ and $T=\infty$ are
\be
\beta_H(T_0)={1\over3}+(1+{m_H^2\over3m_W^2}){\Sigma(N^3)\over\beta_G}
+{\cal O}(\beta_G^{-2}), \quad \beta_H(T=\infty)=\beta_H(T_0) -
{m_H^2\over 108\beta_Gm_W^2}.
\la{betahappro}
\ee

Eq.\nr{betahT} takes correctly into account the divergent mass
counterterms of the 3d theory. In addition to these, there will be
finite renormalisations of the coupling constants $g$ and $\lambda$.
Due to infrared divergences getting more and more serious at higher
orders \cite{templeton} these renormalisations -- which will depend on
the lattice size $N$ -- cannot be calculated perturbatively. We shall
observe that the numerical results will follow the "constant physics"
curve \eq\nr{betahT}, with $N$ dependent mass renormalisation but $N$
independent bare values of $g$ and $\lambda$, rather well. However,
some more $N$ dependence, clearly attributable to finite
renormalisations of the coupling constants, will remain. In any case,
all finite size effects will be very difficult to control. For
example, it is not known how to include the constant mode (the $n_i=0$
term in \eq\nr{Sigmalatt}) term in perturbative calculations.

\subsection{Effective potential on the lattice.}
Monte-Carlo lattice simulations should be compared with the
perturbative calculations of the effective potential on the lattice
rather than with continuum expressions. To get a lattice
generalization for the effective potential one can just change the
integration over momenta in \eq\nr{1looppot} to a finite sum over the
discrete momenta $p_i=(2\pi/aN)n_i, n_i=0,...,N-1$. Including only the
$A_i$ loop term, which is relevant for the first order nature of the
transition, the lattice effective potential becomes
\be
V_{\rm{latt}}=\fr12\gamma(T^2-T_0^2)\phi^2 +\fr14\lambda\phi^4+
{3T\over (aN)^3}\sum_{n_i=0}^{N-1} \biggl[
\log\biggl(1+{(ga\phi/4)^2 \over d}\biggr) -{(ga\phi/4)^2 \over d}\biggr],
\la{vlatt}
\ee
where
\be
d =\sin^2(\pi n_1/N)+\sin^2(\pi n_2/N)+\sin^2(\pi n_3/N).
\ee
Note that in this sum the term with $n_i=0$ must not be included not
only since we cannot handle it but also by definition since the
effective potential is constructed for $x$-independent $\phi$ and thus
constant $\phi$ configurations must not be integrated over.

To study the order of the transition it is convenient to study the
zeroes of
\be
{dV_{\rm{latt}} \over \phi d\phi}= \gamma(T^2-T_0^2)+\lambda\phi^2
-{3g^4\over128}\phi^2 aT{1\over N^3} \sum_{n_i=0}^{N-1}
{1\over d[d+(ga\phi/4)^2]}, \la{vlattprime}
\ee
In the continuum limit $a\to0,N\to\infty,aN\to\infty$ the last term
becomes $-(3/16\pi) g^3T\phi$. For $T=T_0$ one then is solving
$\lambda\phi^2 -(3/16\pi) g^3T\phi=0$, which trivially leads to a
second minimum. This also exists for some temperatures above $T_0$, up
to $T=T_+$. The case of finite lattices is completely different. It is
clear from the lattice expression for the effective potential that it
is an analytic function of $\phi^2$ at least for $\phi^2 < (16/g^2
a^2)\sin^2(\pi/N)$ and , therefore, no term $\sim -\phi$ can appear.
Nevertheless, a second minimum can exist for sufficiently large
lattice sizes. The condition for this simply is that the equation
\be
\lambda={3g^4\over128}aT{1\over N^3} \sum_{n_i=0}^{N-1}
{1\over d[d+(ga\phi/4)^2]} \la{cond}
\ee
have a solution. The right hand side decreases monotonically when
$\phi$ increases and we can take $\phi=0$; if a solution appears it
appears first at $\phi=0$. Inserting $aT=4/(g^2\beta_G)$ and
$\lambda=g^2m_H^2/(8m_W^2)$ gives the relation
\be
\fr43\beta_G{m_H^2\over m_W^2} = {1\over N^3}\sum_{n_i=0}^{N-1}
{1\over d^2} \approx 0.17N. \la{cond1}
\ee
The approximation is the result of an explicit numerical calculation
of the sum.  For the value $\beta_G=20$ used in our numerical
simulations one sees that for $m_H=35$ GeV a first order transition
appears for $N>29$ and for $m_H=80$ GeV for $N>160$.  The use of so
large lattice volumes is not possible in practice, and we at this
stage confined ourselves to smaller $N$, $N \le 20$. If perturbation
theory works well, one should not find any signal of a first order
phase transition in the Monte-Carlo simulations with these small
lattices.

It may be of interest to note that in the limit of large lattices
($a$ = constant, $N  \rightarrow \infty$) the last term in
\eq\nr{vlatt} can be written in the form \cite{elze}
\be
{3T\over a^3}\int_0^\infty dy\,\biggl[{1\over y}(1-\exp[-(ga\phi/4)^2y])
-(ga\phi/4)^2\biggr]\exp(-3y)I_0^3(y), \la{ninfty}
\ee
with the aid of the Bessel function $I_0$.
Numerically this is very close to the continuum result, in particular,
the $\phi\to 0$ limits are the same. In
other words, the critical temperature as well as the metastability range
given in Table 1 are practically the same for very large $N$ and
for the continuum.

\section{Lattice simulations}
Our choice of parameters for the simulations is motivated as follows.
We have already fixed that $g=2/3$ and $m_W=80.6$ GeV.  Since our aim
is to test the validity of perturbation theory we choose $m_H=$ 35
GeV, which makes $\lambda$ small = 0.0105 but is not too close to the
vacuum stability limits of somewhat less than 10 GeV.  For comparison
we also choose $m_H=$ 80 GeV ($\lambda$=0.0547).  We also do not
expect fermions (except perhaps for the top quark) to qualitatively
change the nature of the transition and thus choose $N_F=0$. Light
fermions could simply be included by changing the numerical value of
the coefficients as in \eq\nr{3daction}.  Finally, we choose $\beta_G$
= 20, which basically fixes the lattice spacing $a$ in physical units
via the first equation in \eq\nr{betas}: $a=0.45/T$. As should, this
is smaller than the thermal distance scale $1/T$, the average distance
between particles. Equivalently, in order to describe correctly
effects associated with magnetic sector of the 4-dimensional theory
one must have $a \ll 1/m_M$. With the use of $C_M \approx 2$ and
$\beta_G = 20$ we get $a\approx 0.5/m_M$. For this $a$ the
perturbative Higgs field correlation lengths in Table 1 are 27$a$
($m_H=35$) and 42$a$ ($m_H=80$) while the gauge field correlation
lengths are 6a ($m_H$=35) and 14a ($m_H=80$).  These are so large that
it is realistically not possible to fit an interface between two
phases (broken and unbroken) in the lattice.

\subsection{The update algorithm.}
Because the lattice system described by the action
\eq\nr{latticeaction} has several qualitatively different components,
we used a wide mixture of update algorithms to obtain good
performance.  The SU(2) gauge field was updated as follows: first, we
combined the plaquette action (first term in \eq\nr{latticeaction})
and the $\Phi$-field hopping term (fifth term) to form a local SU(2)
action of the form $\beta^\rmi{eff}_i(x)\tr X_i(x)U_i(x)$,
$X\in\mbox{SU(2)}$.  Using this action, new link matrices were
generated with the Kennedy-Pendleton heat bath method~\cite{kenpen}.
The adjoint field hopping action $S_\rmi{hopp}(U,A_0)$ (second term),
which is quadratic in $U_i$, was then taken into account by accepting
or rejecting the new link matrices with the Metropolis method -- that
is, with the condition $\exp [S_\rmi{hopp}(U^\rmi{new},A_0) -
S_\rmi{hopp}(U^\rmi{old},A_0) ] > r$, where $r$ is a random number
from an uniform distribution between 0 and 1.  The acceptance rate for
the Kennedy-Pendleton heat bath was $\sim 99.5\%$ and for the
accept/reject step $\sim 95\%$.

The length of the adjoint field $R_A = (A_0^a A_0^a)^{1/2}$ was
updated with the Metropolis method, while the colour space direction,
which appears only in the hopping term, was updated with an SO(3) heat
bath.  Similarly, the fundamental Higgs field was divided into radial
and SU(2) parts: $\Phi = RV$, $R>0$, $V \in \mbox{SU(2)}$.  The
$V$-field, which appears only in the hopping term (the fifth term in
\eq\nr{latticeaction}), was updated with the Kennedy-Pendleton heat
bath algorithm, and the length $R$ was updated with the Metropolis
algorithm.

The evolution of the gauge and $A_0$-fields in the simulation time was
very rapid compared to the evolution of the $\Phi$-field.  It is
crucial to make the Higgs field update as effective as possible.
Because of the large $\beta_G$ the gauge background of the Higgs field
is very flat, and it is plausible that one could construct an
effective multigrid or cluster update algorithm.

\subsection{The results.}
Results of the simulations are shown in Figs. 2-6. In Fig.2 we present
the probability distribution of the order parameter $L = \tr
V^\dagger(\bfx)U_i(\bfx)V(\bfx+i)$ for $m_H = 35$ GeV and different
lattice sizes. The two peak structure characteristic of a first order
phase transition is clearly seen on these $8^3$ and $20^3$ lattices,
as well as on intermediate $12^3$ and $16^3$ lattices (not shown). The
change of the curves with the lattice volume is qualitatively
consistent with what one would expect from a first order transition,
namely the two peak structure is more distinguished on the large
lattices. Note that the peak positions do not depend on the lattice
size which indicates that the finite volume effects are not
substantial.

The first order nature of the phase transition can be also seen in
Fig. 3 where the simulation time evolution of the order parameter $L$
is shown.  Here initially the system was confined in the unbroken
phase with small value of $L$, then it jumps to the phase with broken
symmetry and stays there quite a long time. These jumps then continue
in Monte Carlo "time".

An important criterion for the order of the phase transition is the
finite size scaling of the second moment of the order parameter $L$
\cite{borgs}. This is shown on Fig.~4 for various lattice sizes.  The
continuous curves were obtained by combining the individual runs
(performed at various $\beta_H$'s) with the multiple histogram
(Ferrenberg-Swendsen) method \cite{ferrenberg}.  We observe that the
second moment of $L$, as a function of $\beta_H$, develops a narrow
peak as the volume is increased.  The height and the location of the
peak have well-defined infinite volume limits; this behaviour
is characteristic for systems exhibiting a first order phase transition.

Analogous data for a more heavy Higgs ($m_H = 80$ GeV) is presented on
Figs. 5-6. In this case the correlation lengths for the Higgs and $W$
bosons are larger than the lattice size and strong volume dependence
is observed (Fig. 6). No definite conclusion can be made for this case
with the present lattice sizes.

According to Fig.2 the transition for $m_H=35$ GeV takes place at
about $\beta_H=0.34010$ roughly independent of the lattice size. We
thus do not observe precise scaling according to \eq\nr{betahT}, which
would demand that the value corresponding to some fixed temperature
change according to \eq\nr{betahappro} with lattice size $N$. We
ascribe this to the nonperturbative renormalisations of $g$ and
$\lambda$ discussed above.  If we extrapolate to the curve
corresponding to $N=\infty$, we see that $\beta_H=0.34010$ corresponds
to $T_c\approx 85$ GeV, somewhat but not much below the continuum
perturbative value of $T_c=95$ GeV. For $m_H=80$ GeV the results are
less conclusive, but for the largest lattice studied the transition
takes place at $\beta_H=0.3418$. The $m_H$ dependence of
\eq\nr{betahappro} is rather well reproduced, but so far it is
impossible to make a definite conclusion concerning $N$ dependence.
Anyway it is suggested by Fig.1 that again the $T_c$ observed is less
than the perturbative value of 184 GeV.

\section{Conclusions}

We have performed a combined analytical and numerical study of the
finite $T$ electroweak phase transition. First those degrees of
freedom which -- with reasonable degree of confidence -- can be
treated perturbatively were analytically integrated over and an
effective action in the remaining degrees of freedom was derived. This
effective theory is a $T=0$ 3d SU(2) gauge field + adjoint Higgs +
fundamental Higgs theory with known coefficients depending on $T$ and,
quite essentially, on the cutoff of the theory. This bare effective
theory was then latticized with $1/a$, $a$ = lattice spacing, as the
cutoff and the nonperturbative degrees of freedom were treated
numerically with Monte Carlo techniques.

In this first numerical application our aim was to study the structure
of the theory and a fairly small Higgs mass, $m_H=$ 35 GeV was mainly
discussed, with some results given also for $m_H=80$ GeV. A first
order transition was clearly seen at least for the smaller Higgs mass
and the numerical value of $T_c$ is rather close to (but less than)
the value obtained from 1-loop perturbation theory. For this numerical
agreement the computable cutoff dependence of the bare effective
theory was crucial.  With increasing $m_H$ the various characteristic
lengths in the problem increase rapidly and the problem becomes
numerically more and more difficult.

Our lattice Monte Carlo results thus clearly indicate that there is a
first order phase transition, at least for small values of the Higgs
mass. A very intriguing part of the result is that we see a {\em first
order} phase transition also on {\em small} lattices where,
according to the perturbative calculations in Section 4.2, the phase
transition must be of the second order. A number of tests including
the lattice volume dependence of the order parameter, temperature
metastability range, etc. indicate that the first order character of
the phase transition is not a lattice artifact.  These results
indicate that so-called $\phi^3$ term is {\em not} the only source for
the first order character of the electroweak phase transition and that
non-perturbative effects are important as well.  The physical nature
of these non-perturbative effects should be presumably related to the
confinement in 3-dimensional gauge theory.  Unfortunately, the
complete solution of the problem of the electroweak phase transition
by non-perturbative lattice methods requires huge lattices.

\subsubsection*{Acknowledgment}

The authors are grateful to K. Farakos for the collaboration at the
initial stage of this work and providing us his code for 3-dimensional
simulations of 3-dimensional gauge theory with a Higgs doublet.

\newpage
\begin{figure}
\hskip3cm\epsfig{file=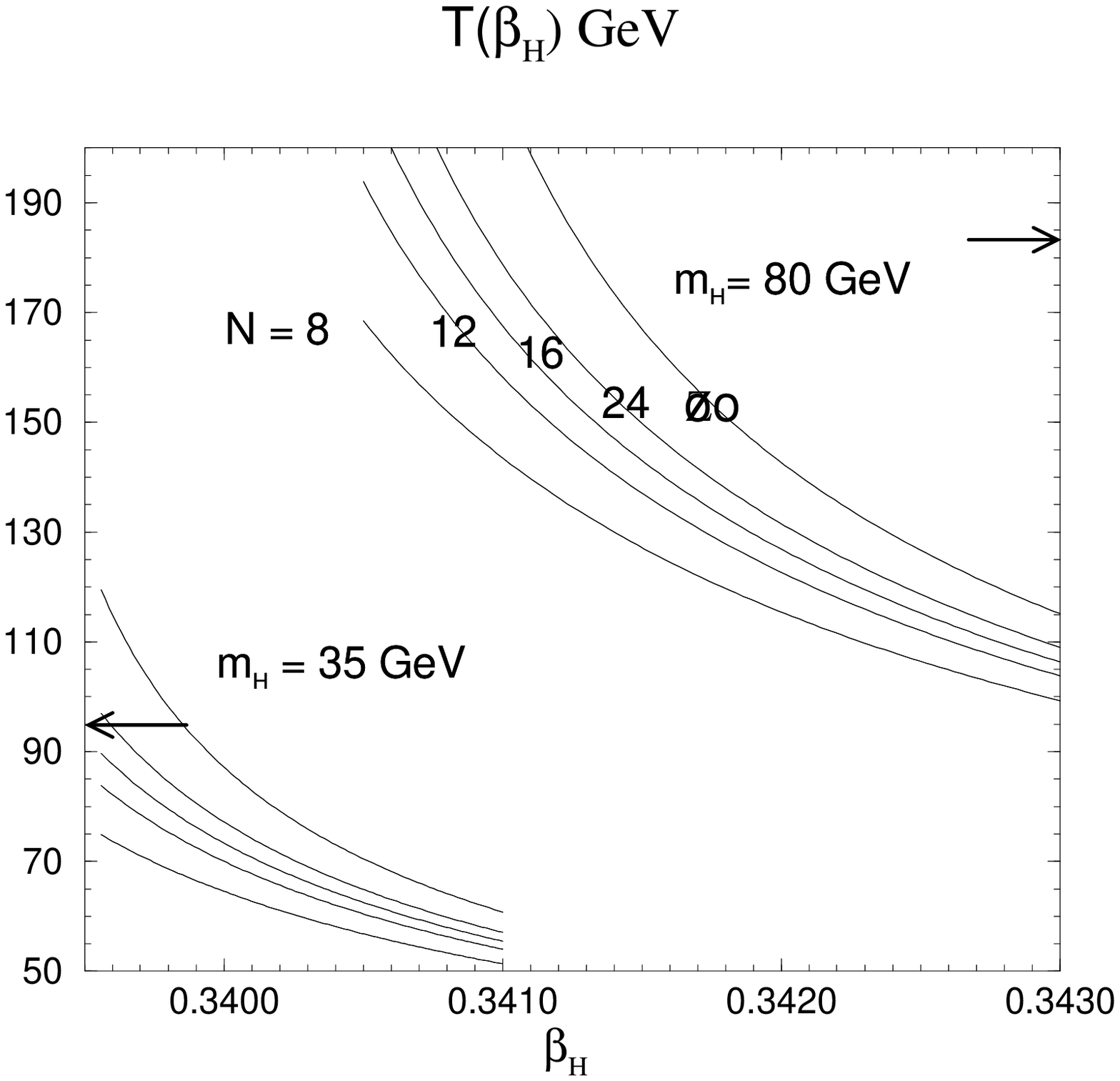,height=8cm}
\vspace{0.5cm}
\caption[0]{The relation between $T$ and $\beta_H$ as given by
\eq\nr{betahT}
for $m_H=$ 35 and 80 GeV, $\beta_G=20$ and $N_F=0$. The arrows
denote the perturbative values of $T_c$ as given in Table 1.}
\label{fig1}
\end{figure}

\begin{figure}
\hskip3cm\epsfig{file=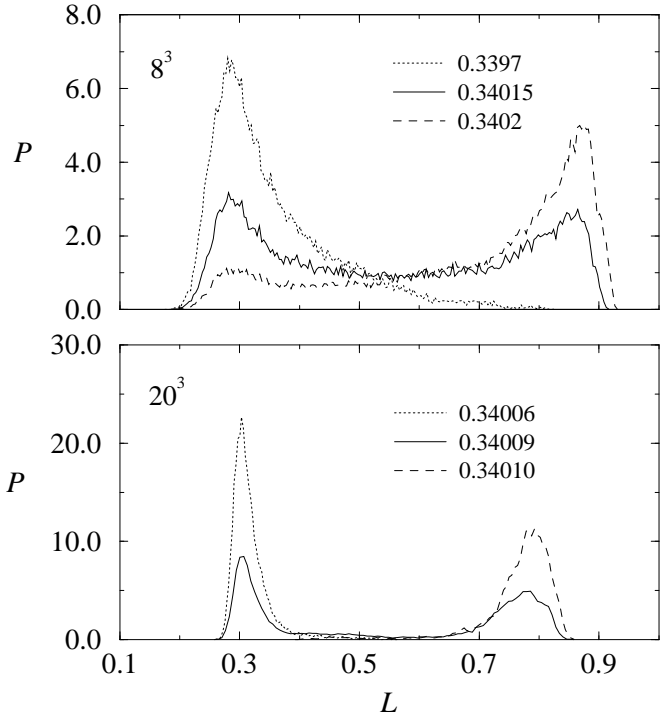,height=8cm}
\vspace{0.5cm}
\caption[0]{The distribution of $L = \tr  V^\dagger(\bfx)U_i(\bfx)V(\bfx+i)$
  for $m_H=35$ GeV for $8^3$ and $20^3$ lattices for $\beta_G = 20$
  and different values of $\beta_H$.}
\label{fig2}
\end{figure}

\begin{figure}
  \hskip3cm\epsfig{file=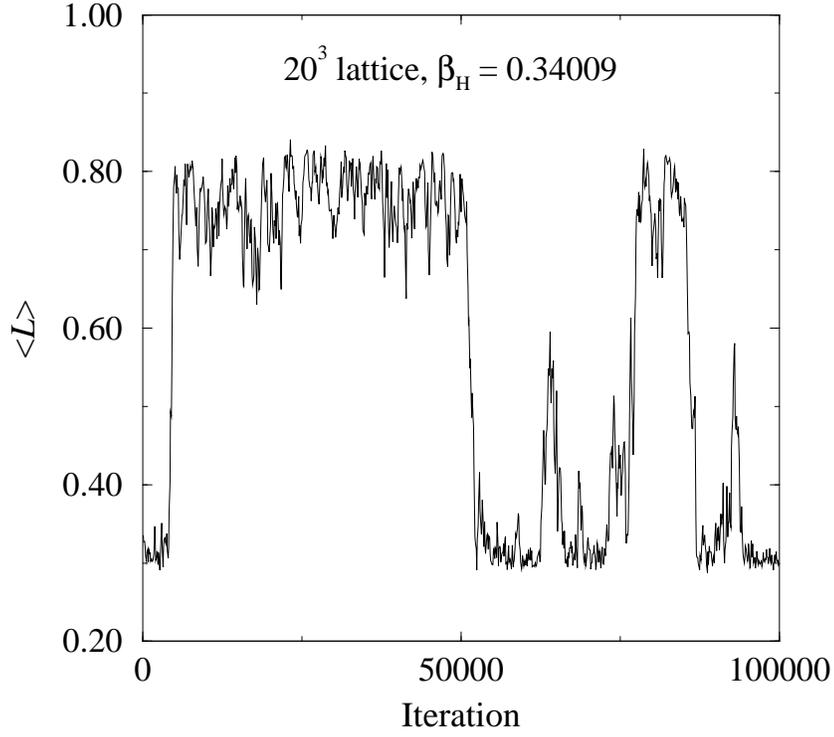,height=8cm}
\vspace{0.5cm}
\caption[0]{The Monte Carlo time history of the lattice simulations for
  $m_H$ = 35 GeV on a $20^3$ lattice near critical $\beta_H$.}
\label{fig3}
\end{figure}

\begin{figure}
  \hskip3cm\epsfig{file=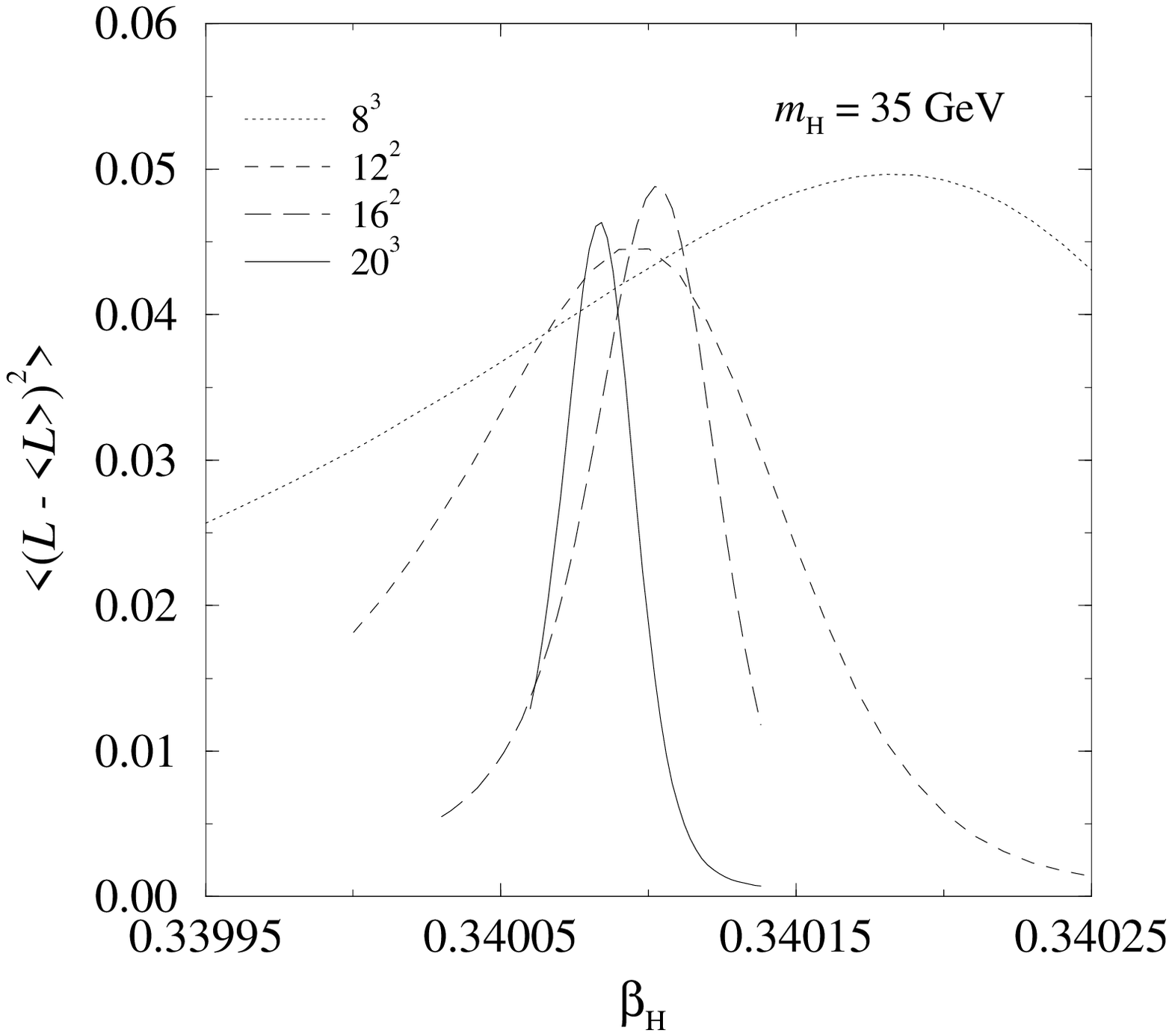,height=8cm}\vspace{0.5cm}
\caption[0]{The dependence of the second moment of the order parameter $L$
  on  $\beta_H$ for different lattice sizes and $m_H = 35$ GeV.}
\label{fig4}
\end{figure}

\begin{figure}
  \hskip3cm\epsfig{file=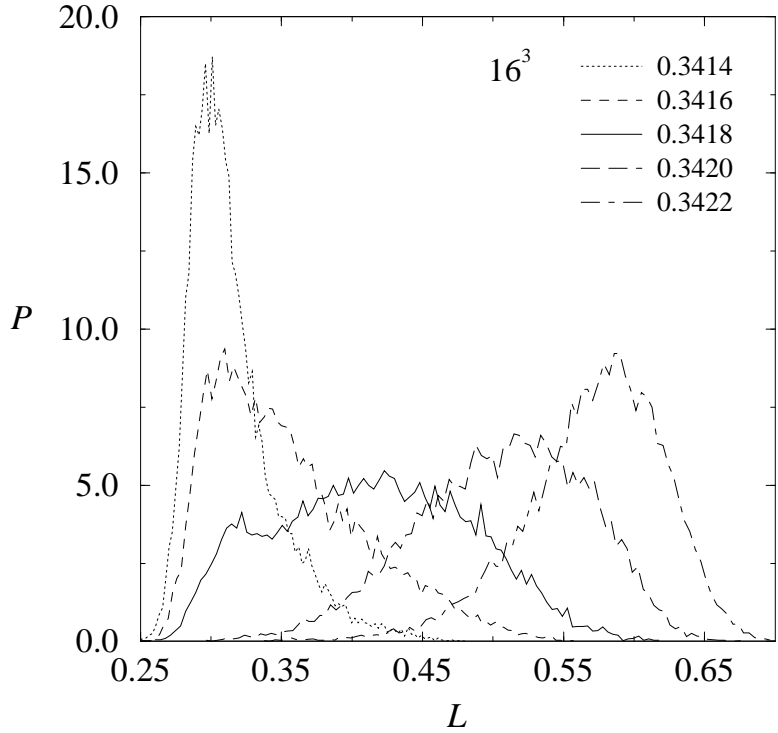,height=8cm}\vspace{0.5cm}
\caption[0]{The distribution of $L = \tr  V^\dagger(\bfx)U_i(\bfx)V(\bfx+i)$
for $m_H=80$ GeV and a $16^3$ lattice for different values of $\beta_H$.}
\label{fig5}
\end{figure}

\begin{figure}
\hskip3cm\epsfig{file=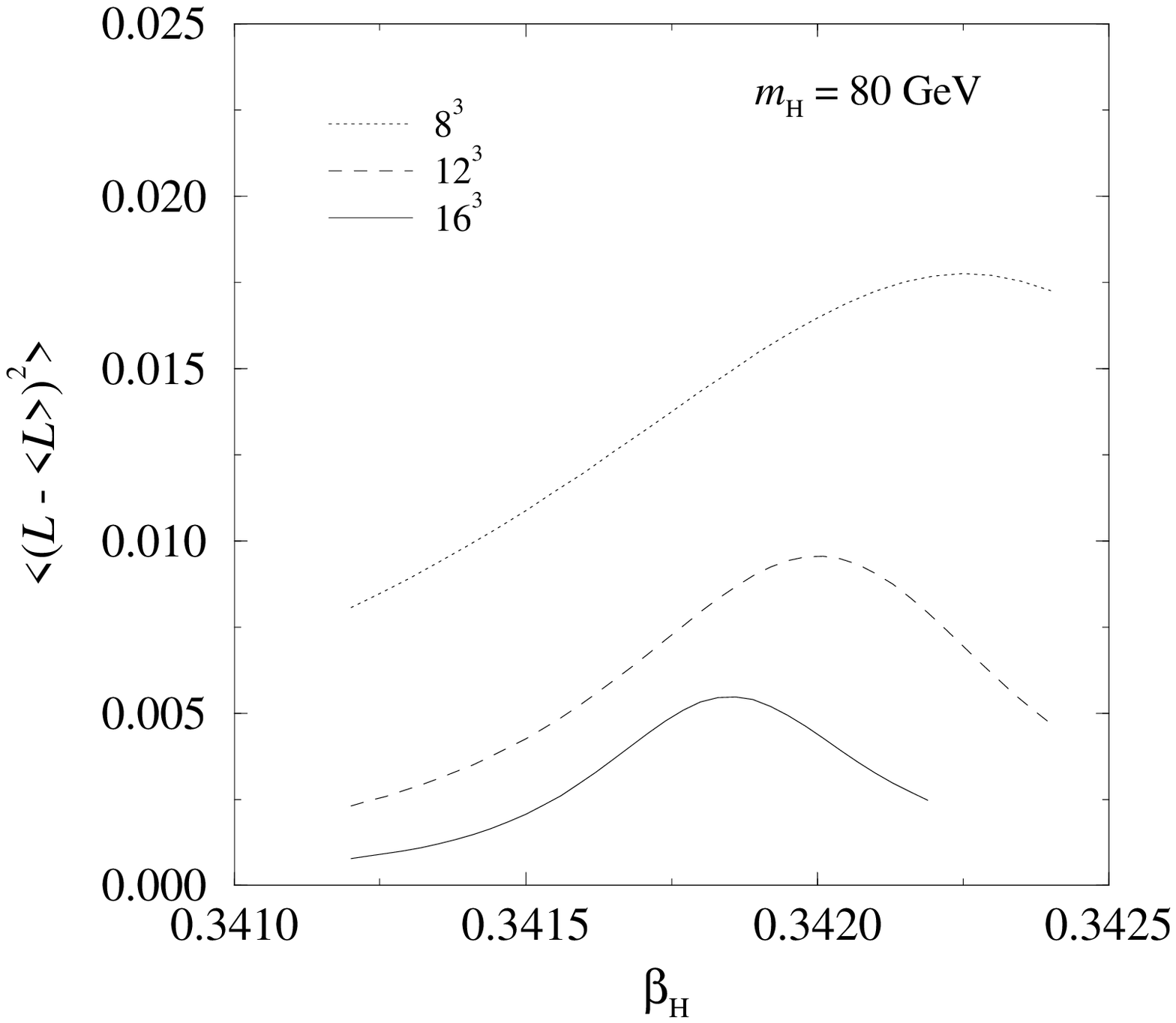,height=8cm}\vspace{0.5cm}
\caption[0]{The dependence of the second moment of the order parameter
$L$ on $\beta_H$ for different lattice sizes and $m_H = 80$ GeV.}
\label{fig6}
\end{figure}


\begin{thebibliography}{99}

\bibitem{kir} D. A. Kirzhnitz, JETP Lett. 15 (1972) 529;\\
D. A. Kirzhnitz and A. D. Linde, Phys. Lett. 72B (1972) 471

\bibitem{s:rev}
M.~Shaposhnikov, In: 1991 Summer School in High Energy
Physics and Cosmology, v. 1, p. 338, World Scientific, 1992;\\
N.Turok, Preprint IMPERIAL-TP-91-92-33, 1992;\\
D.B.Kaplan, A.G.Cohen and A.E.Nelson, Preprint
UCSD-PTH-93-02/BUHEP-93-4, 1993

\bibitem{far} G.~R.~Farrar and M.~E.~Shaposhnikov, Preprint
CERN-TH.6734/RU-93-11, 1993

\bibitem{linde} D.A.Kirzhnitz and A.D.Linde, Ann. Phys. 101 (1976)
195;\\ A.D.Linde, Nucl. Phys. B216 (1983) 421, Rep. Prog. Phys. 47
(1984)925

\bibitem{takahashi} K. Takahashi, Phys. Rev. Lett. 56 (1986) 7

\bibitem{s:sm87} M.~E. Shaposhnikov. Nucl. Phys. B287 (1987) 757; A.
I. Bochkarev and M. E. Shaposhnikov, Mod. Phys. Lett. 2A (1987) 417

\bibitem{anderson} G. W. Anderson and L. J. Hall, Phys. Rev. D45
(1992) 2685 \bibitem{carrington} M. Carrington, Phys. Rev. D45 (1992)
2933

\bibitem{lindeandson} M. Dine, R.G. Leigh, P. Huet, A. Linde and D.
Linde, Phys. Rev. D46 (1992) 550

\bibitem{enqvist} K. Enqvist, J. Ignatius, K. Kajantie and K.
Rummukainen, Phys. Rev. D45 (1992) 3415

\bibitem{brahm}
D.E.Brahm, C.G.Boyd and S.D.H Hsu. Preprint  EFI-92-22, 1992

\bibitem{ae} P.Arnold and E.Espinosa, Preprint UW/PT-92-18, 1992

\bibitem{eqz} M.Quiros, J.R.Espinosa and F.~Zwirner, Preprint
CERN-TH.6577/92, 1992

\bibitem{buch} W. Buchm\"uller, Z. Fodor, T. Helbig and D. Walliser,
Preprint DESY 93-021, 1993

\bibitem{linde:pl80} A.D. Linde, Phys. Lett. 96B (1980) 289;\\
D.~Gross, R.~Pisarski and L.~Yaffe, Rev. Mod. Phys. 53 (1981) 43

\bibitem{kks:magn}A.~Irb\"ack and C.~Peterson, Phys. Lett. 174B (1986)
99;\\ G.~Koutsoumbas, K.~Farakos and S.~Sarantakos, Phys. Lett. 189B
(1986) 173

\bibitem{heller} P. H. Damgaard and U. M. Heller, Phys. Lett. B171
(1986) 442

\bibitem{jersak} H.G.Evertz, J. Jers\'ak and K. Kanaya, Nucl. Phys.
B285 (1987) 229

\bibitem{bunk} B. Bunk, E.-M. Ilgenfritz, J. Kripfganz, A. Schiller,
Phys.  Lett. B284 (1992) 371; Bielefeld Preprint BI-TP 92/46

\bibitem{nadkarni} S. Nadkarni, Phys. Rev. D27 (1983) 917; Phys. Rev.
D38 (1988) 3287

\bibitem{ginsparg} P. Ginsparg, Nucl. Phys. B170 (1980) 388

\bibitem{landsman} N. P. Landsman, Nucl. Phys. B322 (1989) 498

\bibitem{reisz1} T. Reisz, Z. Phys. C53 (1992) 169

\bibitem{lacock} P. Lacock, D. Miller and T. Reisz, Nucl. Phys. B369
(1992) 501

\bibitem{onelooptop}N. V. Krasnikov, Yad. Fiz. 28 (1978) 549;\\
 H. D. Politzer and S. Wolfram, Phys. Lett. 82B (1979) 242

\bibitem{parisi} G. Parisi, Statistical Field Theory, Addison Wesley
1988,
Ch. 5

\bibitem{templeton} R. Jackiw and S. Templeton, Phys. Rev. D23 (1981)
2291

\bibitem{kenpen} A. D. Kennedy and B. J. Pendleton, Phys. Lett. 155B
(1985) 393

\bibitem{elze} H.-T. Elze, K. Kajantie and J. Kapusta, Nucl. Phys.
B304 (1988) 832

\bibitem{borgs} C. Borgs, R. Koteck\'y and S. Miracle-Sole, J. Stat. Phys.
62 (1991) 529

\bibitem{ferrenberg} A. M. Ferrenberg and R. H. Swendsen, Phys. Rev.
Lett. 61 (1988) 2635

\end{thebibliography}
\end{document}